\documentclass[12pt]{amsart}

\usepackage{bbm}
\usepackage{mathrsfs}
\usepackage{amsfonts}
\usepackage{amssymb}
\usepackage[centertags]{amsmath}
\usepackage{amsthm}
\usepackage{graphicx,caption,subcaption}
\usepackage[usenames]{color}

\theoremstyle{plain}
\newtheorem{theorem}{Theorem}[section]

\newtheorem{algorithm}[theorem]{Algorithm}
\newtheorem{prototype}[theorem]{Prototype Theorem}

\theoremstyle{definition}

\theoremstyle{remark}
\newtheorem{remark}{Remark}[section]
\numberwithin{equation}{section}

\newcommand{\ep}{\varepsilon}
\newcommand{\integer}{{\mathbb Z}}
\newcommand{\torus}{{\mathbb T}}
\newcommand{\real}{{\mathbb R}}
\newcommand{\Tau}{{\mathcal T}}

\begin{document}
\title[Analyticity Breakdown in Quasi-periodic Media]
{The Analyticity Breakdown for Frenkel-Kontorova 
Models in Quasi-periodic Media: Numerical Explorations}
\author[T. Blass]{Timothy Blass}
\address{
Center for Nonlinear Analysis\\
Department of Mathematical Sciences\\
Carnegie Mellon University\\
Pittsburgh, PA 15213 (phone: (412) 345-1533, fax: (412) 268-6380)
}
\email{tblass@andrew.cmu.edu}
\author[R. de la Llave]{Rafael de la Llave}
\address{School of Mathematics\\
Georgia Institute of Technology\\
686 Cherry St. \\ 
Atlanta GA 30333-0160
}
\email{rafael.delallave@math.gatech.edu}
\thanks{R.L. has been partially supported by 
NSF grant DMS 1162544. T.B. was supported by the NSF under the
PIRE Grant No. OISE-0967140.}

\maketitle
\begin{abstract} 
We study numerically the \emph{``analyticity breakdown''} 
transition in 1-dimensional  quasi-periodic media. 
This transition corresponds physically to the transition between 
pinned down and sliding ground states. 
Mathematically, it corresponds to the solutions of a 
functional equation losing their analyticity properties. 

We implemented some recent numerical algorithms that are 
efficient and backed up by rigorous results so that we can 
compute with confidence even close to the breakdown. 

We have uncovered several phenomena that we believe deserve 
a theoretical explanation: 
A) The transition happens in a smooth surface. 
B) There are scaling relations near breakdown. 
C) The scaling near breakdown is very anisotropic. 
Derivatives in different directions blow up at different rates. 

Similar phenomena seem to happen in other KAM problems. 
\end{abstract}



\section{Introduction} 
\label{sec:intro} 
In this paper we present detailed numerical results on 
the \emph{analyticity breakdown} transition in models of 
one-dimensional quasi-periodic media. 
This analyticity breakdown transition is widely accepted to correspond 
to the transition between pinned down and sliding ground states. Hence, 
the position of the transition affects the properties of 
the material described by the models. For the physical 
motivation of the theory see Section~\ref{sec:physics} and 
\cite{AubryA80,vanErpthesis}.

{From} the mathematical point of view, the problem is 
to study the analyticity properties of the solution of 
a functional equation depending on parameters (the function is 
the hull function parameterizing a ground state and the functional 
equation is the expression that the system is in equilibrium). 
It is expected that, for some parameter values, the solution is 
analytic but that, as we move parameters, the domain of analyticity 
decreases and eventually disappears.  See Section~\ref{sec:physics} for 
a more precise description of the problem. 

The phenomenon of analyticity breakdown is very similar to the 
widely studied phenomena of breakdown of KAM tori in mechanics. 
Indeed, the problem of analyticity breakdown in periodic media
is completely equivalent to the breakdown of KAM circles in 
two dimensional twist mappings. In the case of quasi-periodic 
media considered here, the problem does not seem to have an 
easy dynamical representation and the treatment of the 
functional equations describing the equilibria for KAM theory is 
very different from the Hamiltonian counterpart because, 
besides the lack of a dynamical interpretation, the system 
has an extra frequency.  The KAM theory for the equilibrium 
equation in  quasi-periodic 
media has been obtained recently in 
\cite{SuL12a, SuL12b}.  

The KAM theory in \cite{SuL12a, SuL12b} is very different from 
the regular KAM theory for two reasons: one is 
that the models do not admit an easy dynamical interpretation 
and a second reason is 
because the quasi-periodicity of the substratum introduces the 
an extra variable. 
This extra frequency has very drastic effects, which are not just 
mathematical artifacts. For instance, 
there are counterexamples to some standard predictions of 
the Aubry-Mather theory which describes the ground states in periodic 
media \cite{LionsS02}. In the periodic case, when Aubry-Mather theory 
applies, the analyticity breakdown transition could be described 
as saying that the smooth tori break up into Cantori 
(see \cite{Percival}). In the case of quasi-periodic media, we 
are not sure of what could take the place of the Aubry-Mather theory after 
the breakdown of KAM theory. This is a question that deserves further 
study, but which 
we are not considering in this paper.

Nevertheless, we expect that some of the phenomena studied in detail 
here, notably the  anisotropic regularity at breakdown happen also 
in other KAM  problems involving two-dimensional functions. 
Indications to that effect have already been found qualitatively 
\cite{Tompaidis92a, HaroS97,CellettiFL04}, but in this paper we 
can obtain rather quantitative aspects of the phenomenon. 

Since the main goal is to compute accurately functions as close 
as possible to their breakdown, it is clear that the numerics
are going to be delicate and that it is very important to have criteria
that allows to be confident that the computed solutions are 
correct. We note that the experience shows that close to the breakdown, 
the truncated equations admit many \emph{spurious} solutions 
that do not correspond to truncations of true solutions of 
the equations \cite{VanErp'02, GentileE05}. 

In this paper we have implemented the algorithms suggested in 
\cite{SuL12a}. The main results of \cite{SuL12a} are based on 
a Newton method which is very efficient (see Section~\ref{sec:algorithm}). 
If the function is discretized using $N$ Fourier coefficients, 
the Algorithm \ref{alg:Newton} gives quadratic convergence 
but requires only $O(N)$ storage and $O(N \log (N))$ arithmetic operations. 
Note that, even if we obtain the Newton-like quadratic convergence 
a step does not require to store (and much less to invert) an
$N \times N$ matrix. 

Furthermore, the main result of  \cite{SuL12a} is a result in 
\emph{a posteriori format}. It asserts that, if 
we have a function which solves approximately the invariance 
equation (e.g. a numerical solution of the equation, where the
error of truncation and round off is small) and that satisfy 
some easy to verify non-degeneracy assumptions corresponds 
to a true solution of the problem. Hence, by computing with 
a truncation, estimating the truncation error by changing the degree
of the truncation
and monitoring the non-degeneracy conditions, we may be reasonably 
confident that we have obtained a true solution. 

Of course, a numerical implementation requires many more details 
than a mathematical algorithm and one has to discuss implementation issues
such as data structures used, choice of iteration 
steps, order of truncation, etc. 
These considerations are undertaken in 
Section~\ref{sec:numerics}.  The results are presented in 
Section~\ref{sec:results} and the conclusions are presented in 
Section~\ref{sec:conclusions}. 
We anticipate that the conclusions are 
A) The breakdown happens in a smooth surface, 
B) There are scaling relations. Sobolev norms of 
the function blow up as a power of the distance to 
the critical surface. 
C) The breakdown is extremely anisotropic. 

Of course, A) and B) have been observed many times in phase transitions
and are one of the predictions of renormalization group theory
but C) seems somewhat unexpected. There have been visual observations 
indicating that the tori near breakdown look \emph{``filamented''}. 
We think that there are reasons to expect that this phenomena will 
happen in other KAM problems involving several variables and we hope 
to come to that problem. In this paper, we will just present some 
numerical observations in a class of models. 

\section{The problem of analyticity breakdown} 
\label{sec:physics} 
The simplest way to motivate the models studied here is 
to consider the problem of deposition of a material over 
a one-dimensional quasi-periodic substratum. 

If $u_i$ represents the position of the $i^{\rm th}$ particle
of the deposited material, the state of the system is 
given by the configuration $\{ u_i\}_{i \in \integer}$, which is 
sequence giving the position of all the deposited particles. 
Note that $u_i$ will be a  real variable. 

The Physics of the problem is obtained by assigning an formal energy 
to the configurations \cite{Ruelle,Israel79,Mattis}. 
This formal energy is obtained as a formal 
sum of the energies associated to every finite set of particles. 
Hopefully, most of the terms will be zero or decrease very fast 
with the diameter. Nevertheless, due to the translation invariance 
one does not hope that the sum of the energies converges. 

The main example considered in 
this paper will be  the quasi-periodic Frenkel-Kontorova model given 
by: 
\begin{equation} \label{QPFK} 
S(\{u\}_{i\in \integer})=
\sum_{n\in\mathbb{Z}}\frac{1}{2}(u_n-u_{n+1})^2-\hat V(\alpha u_n) 
\end{equation} 
where $\hat V: \torus^d \rightarrow \real$ and 
$\alpha \in \real^d$ is a sufficiently irrational vector. 

The term $\frac{1}{2}(u_n-u_{n+1})^2$ represents the 
interaction between  nearest neighbors and the 
term $\hat V(\alpha u_n)$ represents the interaction with the substratum 
at the position $u_n$. Note that this function is quasi-periodic
as a reflection of the fact that the substratum  is quasi-periodic. 
The classical Frenkel-Kontorova model 
\cite{FrenkelK39} considered the case of periodic potentials with $d=1$.  

Other physical interpretations of the model are possible, for 
example, the classical \cite{FrenkelK39} introduced
the Frenkel-Kontorova model to describe the motion of planar dislocations 
in a crystal. 

\subsection{Some standard definitions}
Now we recall some standard definitions \cite{AubryA80, AubryD83}. 

A configuration $\{u_i\}_{i \in \integer}$ is in 
equilibrium when
\begin{equation} \label{equilibrium}
\partial_{u_i} S( \{u_i\}_{i \in \integer} ) = 0 
\end{equation}

In the case \eqref{QPFK}, the equilibrium equations 
\eqref{equilibrium} become: 
\begin{equation} \label{QPFKequilibrium} 
\alpha \cdot \nabla \hat V( \alpha u_i) 
+ u_{i+1} + u_{i-1} - 2 u_i = 0 
\end{equation} 

Note that, even if the sum in \eqref{QPFK} is a formal sum, 
the equilibrium equations \eqref{QPFKequilibrium} are very well defined. 

A configuration $\{u_i\}_{i \in \integer}$ is a \emph{ground state} 
when 
\begin{equation} \label{groundstate} 
S(u + \eta) - S(u)  \ge 0
\end{equation} 
for all $\eta$ that vanish at all  but a finite number of
indices.
  Note that the left hand side of \eqref{groundstate} can be given 
a good interpretation in the model \eqref{QPFK} 
 because the two formal sums only differ in a a
finite number of sites. 

A configuration is given by a \emph{hull function} when we
can find $h: \torus^d \rightarrow \real$ and $\omega \in \real$ 
such that 
\begin{equation}\label{hull} 
u_n = \omega n + h( n \omega \alpha) .
\end{equation} 

\subsection{Equilibrium equations for hull functions. 
Formulation of our problem}

Substituting \eqref{hull} into \eqref{equilibrium}, we obtain that  
an equilibrium configuration is given by a hull function if and only 
if 
\begin{equation}\label{hulleq1} 
h( n\omega \alpha + \omega \alpha)  
+h( n\omega \alpha  - \omega \alpha)   
- 2 h(n \omega \alpha)  
+ \left( \alpha \cdot \nabla\right) \hat V( n\omega \alpha + h(n \omega \alpha) ).
\end{equation} 

If $\{n \omega \alpha\}_{n \in \integer}$  is dense on $\torus^d$ 
(which is well known to happen if and only if 
$k \cdot \omega \alpha \notin \integer$ for $k \in \integer^d - \{0\}$, 
\cite{KatokH95}), for a continuous $h$ we obtain that  \eqref{hulleq1} is 
equivalent to
\begin{equation}\label{hullequilibrium}
h( \theta + \omega \alpha)  
+h( \theta  - \omega \alpha)   
- 2 h(\theta)  
+ \left( \alpha \cdot \nabla\right) \hat V( \theta + h(\theta) ).
\end{equation} 

Of course, it could well happen that there are discontinuous 
hull functions, but we are interested in studying precisely 
the existence of continuous -- indeed analytic -- functions. 
 
Hence, \eqref{hullequilibrium} will be the centerpiece of our
attention. We will fix $\omega \alpha$ satisfying good 
number theoretical properties, assume that $\hat V$ depends
on parameters and study numerically the set of parameters for 
which \eqref{hullequilibrium} has an analytic solution. We will 
pay special attention to the behavior of these solutions near the 
boundary of existence. 

We note that the existence of a continuous solution of 
\eqref{hullequilibrium} implies that there is a continuous family of 
equilibrium configurations. On can note that for any $\sigma_0 \in 
\torus^d$, $h_{\sigma_0}$ given 
by 
\begin{equation} \label{gauge}
h_{\sigma_0}(\theta) = h(\theta + \sigma_0)
\end{equation} 
is also 
a solution of \eqref{hullequilibrium}.  Hence, the 
configuration corresponding to it according to 
\eqref{hull} is also an equilibrium configuration. 
The fact that we obtain a continuous family of configurations which are 
equilibria indicates that there is no energetic impediment to 
the solutions making small transitions and sliding. Hence one expects 
that the solutions can be modified easily.  In contrast, if 
the set of equilibrium configurations is discontinuous, there 
may be an energetic  barrier (the Peierls-Nabarro barrier) to perform 
jumps from a configuration to the nearest one and the solutions are 
pinned down. 

Notice that $h_{\sigma_0} $ corresponds to choosing a different origin of
coordinates to the internal phase of the problem. This can be considered 
as a gauge symmetry of the problem. The Ward identities associated 
to this gauge symmetry play an important role in the study in 
\cite{Rafael'08, CallejaL10,SuL12a, SuL12b} and also in our numerical 
treatment.

\section{Mathematical results} 

\subsection{An overview of the rigorous results} 
In the periodic case ($d = 1$), the problem of existence of 
continuous hull functions satisfying \eqref{hullequilibrium} is 
equivalent to the problem of existence of rotational invariant circles for 
a twist mapping \cite{AubryD83}. In the periodic case, there are
several well developed mathematical theories which can lead to an 
understanding of the problem. 
\begin{itemize} 
\item 
Kolmogorov-Arnold-Moser (KAM) theory \cite{Moser66a, Herman83}. 
\item 
Aubry-Mather theory \cite{AubryD83,Mather82,Moser86}. 
\item
Renormalization group theory \cite{MacKay82, Koch}. 
\end{itemize} 
In the periodic case, there are a number of numerical 
methods to study the breakdown. The appendix A of \cite{CallejaL10} 
contains a comparative summary of the literature and  compares the methods. 

The KAM theory is perturbative but produces analytic solutions of 
\eqref{hullequilibrium} when the frequency satisfies 
some number theoretic properties (called  \emph{``Diophantine''}). 
Aubry-Mather theory is non-perturbative and it does not require 
any number theoretic properties of the function. On the other hand, the 
solutions of \eqref{hullequilibrium} may be discontinuous. 
The renormalization theory has as an aim to describe a subset of 
the boundary of analyticity. It was originally developed based 
on non-rigorous arguments supported by very careful numerics
\cite{MacKay82}, but by now there are quite a number of rigorous 
results (some of them proved by computer-assisted proofs). 

In the quasi-periodic case considered here, the situation is very different. 
The Aubry-Mather theory seems somewhat problematic due
to the examples in \cite{LionsS02} (but there are some 
topological results in \cite{Gambaudo}). To the best of 
our knowledge, there is no renormalization group theory for 
general quasi-periodic FK models (see, however \cite{MestelO00}). 
There has been a recent development of a KAM theory in \cite{SuL12a, SuL12b} 
which will be the basis of our work. 

\subsection{A posteriori results} 
The main result of \cite{SuL12a} is formulated in an a-posteriori 
format standard in numerical analysis. 

We define the operator $\Tau$ acting on hull functions by: 
\begin{equation}\label{Tau} 
\Tau[h](\theta) \equiv h( \theta + \omega \alpha)  
+h( \theta  - \omega \alpha)   
- 2 h(\theta)  
+ \left( \alpha \cdot \nabla\right) \hat V( \theta + h(\theta) )
\end{equation} 
So that \eqref{hullequilibrium},  the equilibrium equation for hull 
functions can be concisely expressed as
\begin{equation}
  \label{Tau_zero}
\Tau[h] = 0   
\end{equation}

The main result of \cite{SuL12a} says (omitting some mathematical
assumptions on regularity, definition of norms
 and number theoretic properties of 
$\omega\alpha$)  that if we can find a function that satisfies approximately
the equation (in some norm) and satisfies some non-degeneracy conditions, 
we can find a true solution and bound the distance from 
the true solution to the original guess (measured in another norm) 
by a constant times the size of the residual of the initial approximate 
solution. 

\begin{prototype}\label{aposteriori}
Assume that we can find a function $h_0: \torus^d \rightarrow \real$. 
such that 
\[
\begin{split} 
& || \Tau[ h_0] ||_1 \le \epsilon  \\
& M_1(h_0) \le A_1, \ldots,  M_m(h_0)\le A_m 
\end{split} 
\]
then, if 
\[
\epsilon \le \epsilon^*(A_1,\ldots, A_m)  
\]
there exists a true solution $h^*$ of $\Tau(h^*) = 0$. 
and 
\[ 
||h^* -  h_0||_2 \le K(A_1,\cdots, A_m) \epsilon
\]
Furthermore, the solution is the only solution (up to the gauge 
transformations associated to the change of origin in \eqref{gauge}) 
\end{prototype} 

In the above, $M_1,\ldots, M_m$ are rather explicit condition numbers. 
and $\epsilon^*$ is also an explicit function. 
Note that, as customary in KAM theory, there may be a loss of regularity
and that the norm in which we measure the error may be different than 
the norm in which we reach the conclusions.  In \cite{SuL12a}, 
one can find results where the norms $||\cdot ||_1, || \cdot ||_2$ 
are norms in spaces of analytic functions and also results for 
Sobolev norms. 

In applications, the approximate solution will be the product  of a
numerical calculation. It is important that in view of  the above result, 
that to be confident that the approximate solution corresponds 
to a true solution, we do not need to study the algorithm used 
to produce it. We just need to verify that the condition numbers
are reasonable and that the numerical error is small compared to them. 

As we will see, the method of proof in \cite{SuL12a} 
is based on a rapidly convergent iterative method
which is explicitly described. This method
leads to an efficient algorithm (See Section~\ref{sec:algorithm},
which 
we have implemented in
Section~\ref{sec:numerics} and which is the basis of our results. 
It is also important that the method leads to a numerically accesible criteria
for breakdown, which we discuss in Section~\ref{sec:criterion}. 

\subsection{A numerically accessible criterion for breakdown of 
analyticity} 
\label{sec:criterion} 
In \cite{CallejaL10}, it was shown in great generality that, 
when one has an a posteriori result of the form 
Prototype~\ref{aposteriori}
working for Sobolev and analytic spaces, one can get several results 
automatically. 

{\bf Bootstrap of regularity}
All the solutions which are in a Sobolev space of order $r \ge r_0$
are analytic.

The critical value $r_0$ depends only on the number theoretic properties 
of $\omega \alpha$. 

{\bf Criterion for breakdown of analyticity}
If we consider a family depending on parameters, and obtain a 
family of approximate solutions with bounded non-degeneracy conditions, 
the parameters approach the boundary if and only if Sobolev norms 
of high enough order go to infinity.

Hence, it is clear how one should proceed. Implement the algorithm 
described in Section~\ref{sec:algorithm}, run it monitoring the 
condition numbers (so that we are sure that we 
are not considering any spurious solutions) 
 and the Sobolev norms. If the Sobolev norms blow 
up dramatically and nothing else happens, we are approaching the breakdown. 
Since the algorithm gives the Fourier coefficients of the 
hull function, the computation of the Sobolev norm is rather straightforward. 

It is important to note that this criterion does not require a
dynamical interpretation and in \cite{CallejaL09, CallejaL10} one can find 
a study of systems with long-range interaction for which there is no 
dynamical interpretation.  The paper \cite{CallejaL10} contains a 
comparison with other methods. 

The above criterion for blow up has been implemented several 
times. For twist mappings it has been implemented in 
\cite{CallejaL09, CallejaL10b,FoxM13}. For dissipative systems it 
has been implemented in \cite{CallejaC10,  CallejaF12}.  

\subsection{An efficient algorithm for a Newton method} 
\label{sec:algorithm} 

The proof of \cite{SuL12a} is based on a Newton-like method that given 
an approximate solution produces a much more accurate 
one. The error after the correction is, roughly, the square of 
the error before the correction, but there is a loss of regularity. 
It is well known in the Nash-Moser hard implicit function theorems 
that these methods still lead to convergence. 

The algorithm in \cite{SuL12a}, which we detail below, is obtained 
taking advantage of several identities satisfied by 
\eqref{hullequilibrium} that come from its variational character 
and the gauge symmetry \eqref{gauge}. 

These identities reduce the Newton step to a sequence of elementary
operations on functions (composing, taking derivatives, multiplying, 
solving difference equations with constant coefficients). A remarkable
feature is that all of these operations are diagonal either in 
a discretization in Fourier terms or in  a discretization based
on points on a grid. Of course, the Fast Fourier Transform allows 
to pass from one discretization to another in a easy way. 

The upshot is that if we discretize the function $h$ in terms of 
$N$ Fourier coefficients and the values at $N$ points in a grid, 
a Newton-like step requires $O(N)$ storage and $O(N \log(N) $ operations. 

In practice, the constants in front of
the asymptotics are moderate, since there are very optimized implementations 
of FFT for almost any architecture.  As we will see, it is 
quite possible to use $N \approx 10^5$ in a common desktop. 

For the sake of completeness, we include verbatim the algorithm from 
\cite{SuL12a}. 

The algorithm of \cite{SuL12a} is actually slightly more general 
instead of \eqref{Tau_zero}
it consideres the equation 
\begin{equation}\label{modified} 
\Tau[h] + \lambda = 0 
\end{equation}
where $\lambda \in \mathbb{R}$,  and 
both $h$ and $\lambda$ are the unknowns.

The equation \eqref{modified} allows to consider also 
also forces that do not come
from a potential.  Nevertheless, it is shown
in \cite{SuL12a} that when the forces come from a potential, 
any solution of \eqref{modified} satisfies $\lambda = 0$ and, hence 
gives a solution of \eqref{Tau_zero}.

As we will see, the algorithm consists of 
steps which are all elementary manipulations of 
functions (derivatives, multiplications, compositions) 
as well as solving the so called cohomology equations. 
That is, given $b(\theta)$ periodic of period $1$ 
and with $\int b(\theta) \, d \theta = 0 $, find 
$W$ also periodic with period $1$ and with average $0$ such 
that 
\begin{equation} \label{cohomology} 
W - W \circ T_\sigma = b 
\end{equation}
where $T_\sigma(\theta) = \theta + \sigma$. 

Note that \eqref{cohomology} can be solved very efficietly 
using Fourier coefficients and that it is equivalent to 
the equation for Fourier coefficients 
\[
\hat W_k( 1 - e^{2 \pi i k \cdot \sigma} ) = \hat b_k
\]
The factor in parenthesis in the LHS vanishes for $k = 0$, 
and the equation for $k  = 0$ is overdetermined. It does not have 
a solution if $\hat b_0 \ne 0$, but if $\hat b_0 = 0 $ any
$\hat W_0$ is a solution, we choose the solution with $\hat W_0 = 0$. 
If $\omega $ is irrational, the term in parenthesis 
does not vanish for $k \ne 0$ and, if it is Diophantine, we 
have lower bounds for it that lead to estimates for the 
solutions of the cohomology equation \cite{Russmann75,Russmann76}. 
These estimates are used in \cite{SuL12a}  to prove convergence of the 
algorithm, but in this paper, we will not discuss estimates.

\begin{algorithm} \label{alg:Newton}

Given 
$h:\torus^d\rightarrow \real,~\lambda\in \real$ 
 with $h(\theta) =\sum_{k\in \integer^d}\hat{h}_k
e^{2\pi i k\theta}$ and $\tilde{h}(t)= h(\alpha t)$
for $t \in\real$ and any irrational vector
$\alpha\in\real^d$, we will perform the following calculations
(where $\langle \cdot \rangle$ denotes the average)

\begin{itemize}
\item [1)] Let $\mathscr{L}=\tilde{h}(t +\omega)+\tilde{h}(t -\omega)-2
\tilde{h}(t )$. 

In Fourier components
$\hat{\mathscr{L}}_k=2(\cos{\omega \alpha\cdot k}-1)\hat{h}_k$.

\item [2)] We calculate $\hat{U} \equiv \hat V(\theta + h(\theta)$.

\item [3)] Calculate $e=\mathscr{L}+\hat{U}+\lambda \equiv \Tau[h]$

\item [4)] Calculate $\hat{l}=1+\partial_\alpha\hat{h}$. 

In Fourier components $\hat{l}_k=\delta_{k,0}+2\pi i (k\cdot
\alpha) \hat{h}_k$.

\item [5)] Let $f=\hat{l}\cdot e$.

\item [6)] Choose $\delta=-f_0$.

\item [7)] Denote $b=\hat{l} \cdot (e+\delta)$.

\item [8)] Solve the cohomology equation \eqref{cohomology} for
$\hat{W}^0$ with zero average. That is,
$\hat{W}^0_k=\frac{b_k}{2(\cos{\omega\alpha\cdot k}-1)}$.

\item [9)] Take $\bar{\hat{W}}=-\frac{\left\langle \frac{\hat{W}^0}{\hat{l}\cdot \hat{l}\circ
T_{-\omega\alpha}} \right\rangle}{\left\langle\frac{1}{\hat{l}\cdot\hat{l}\circ
T_{-\omega\alpha}}\right\rangle}.$

\item [10)] Calculate $\hat{W}=\hat{W}^0+\bar{\hat{W}}$.

\item [11)] Solve the  cohomology equation for 
$\tilde{\Delta}$. That is,
$\tilde{\Delta}_k=\frac{a_k}{2(\cos{\omega\alpha\cdot k}-1)}$ where
$a=\frac{\hat{W}}{\hat{l}\cdot \hat{l}\circ T_{-\omega\alpha}}$.

\item [12)] We obtain $\hat{\Delta}=\tilde{\Delta}\cdot \hat{l}$.

\item [13)] The new improved solution is $h + \Delta$, $\lambda + \delta$.

\end{itemize}
\end{algorithm}

\section{Numerical Implementation}
\label{sec:numerics} 
The most computationally expensive steps in Algorithm \ref{alg:Newton}
are diagonal in either real space or in frequency space, so we use the
Fast Fourier Transform multiple times in each application of the
Newton step. We use
Matlab, and typically $2^{18}$ Fourier modes,
so a discretization for $h$ is a $512\times 512$ complex-double. All
non-degeneracy conditions are satisfied if the norm of the error is
small enough, so we monitor the error throughout. 

To implement the algorithm, we compute 
the error of an initial guess, then perform the
Newton step to improve the guess, and then compute the new error for
the improvement. If the new error
is not below a threshold, the Newton step is performed again
and the process repeats. If the new error is below
the threshold, we conclude a torus exists nearby and we 
increase the parameter values.
If the Newton step fails to bring the error below the 
threshold (after some number of attempts)
 or if the
error increases, we end the program and 
start again with a better initial guess. 

We explore the parameter space along rays at a fixed angle $\theta$ by
starting at $(\ep_1,\ep_2) = (0,0)$,
and increasing $(\ep_1,\ep_2) \to (\ep_1 + \Delta \ep, \ep_2 + \Delta \ep \tan \theta)$.
We use the most recently computed invariant
torus as our initial guess (possibly scaled a bit)
at the next parameter value.
If the Newton step cannot bring the error
low enough after a certain number of iterations, we take data from the last point of 
convergence, and try again but with a close parameter value. For example,
if the program converged with $(\ep_1,\ep_2)$ and we take the step 
$\ep_1 \to \ep_1 + \Delta \ep_1, \, \ep_2 \to \ep_2 + \Delta \ep_2$, 
at which point the program fails, then
we try again with $\ep_1 \to \ep_1 + \frac12 \Delta \ep_1, \, 
\ep_2 \to \ep_2 + \frac12 \Delta \ep_2$.
If the program fails to converge, and we have taken a very small step in parameter
space, then we terminate the program and consider 
the lack of convergence as an indication that the 
boundary of analyticity is very close.

The Newton step itself can bring in errors. 
The FFT is used extensively, and occasionally
a mode $\hat{h}_k$ with small modulus will appear from computer error, which
can be amplified as the iterations continue. This can lead to 
problems with convergence due to artificially large Sobolev norms.
 We have found it necessary to monitor the appearance
of false modes and to eliminated them. We implement a method that sets to zero all
modes whose amplitude is below a certain threshold: they are deemed to be numerical
errors and are set to zero so no amplification is possible. 
This slows the convergence
of the Newton method, but allows us to compute tori much closer to the boundary
of analyticity. This is a delicate process; if done
poorly, it can cause the Newton step to return the zero function, 
and thus fail to find an improvement. 
We use a cut-off threshold that is allowed to vary.

When implementing the algorithm, we can only consider a 
finite number of Fourier modes, which
can lead to ``spurious'' solutions. That is, solutions of the discretized problem
that do not correspond to solutions of the original problem. 
For example, when we are looking for a solution to $\Tau[h]=0$, and we 
replace this with a discretized problem $\Tau_N[h]=0$ using $N$ Fourier
modes, we need to ensure the discretized solution $h_N$ 
satisfying $\Tau_N[h_N]=0$
will also satisfy   $\Tau[h_N] \approx 0$.
To avoid computing discrete solutions $h_N$ that satisfy $\Tau_N[h_N]=0$
but do not satisfy $\Tau[h_N]\approx 0$, we compute the error of $h_N$ on a finer grid.
That is, check if $\Tau_{2N}[h_N] \approx 0$. 

We will consider two different models, whose potentials are given by
\begin{align}
  \label{model1}
  & V_1(\phi_1,\phi_2)   = -\frac{\ep_1}{2\pi}\cos(2\pi \phi_1 ) -\frac{\ep_2}{2\pi}
\cos(2\pi \phi_2),\\
& V_2(\phi_1,\phi_2) = -\frac{\ep_1}{4\pi} \cos(4\pi \phi_1 + 4\pi \phi_2)
- \frac{\ep_2}{2\pi}\left(\cos(2\pi \phi_1)  +
\cos(2\pi \phi_2)\right).   \label{model2}
\end{align}

\subsection{Results for Models \eqref{model1} and \eqref{model2}} \label{sec:results}
The results we present 
are for the values
\begin{equation}
  \label{freq_val}
  \omega = 1, \qquad \alpha = (1.246979603717467,\,
2.801937735804838).
\end{equation}
These numbers are roots of cubic polynomials 
($x^3 - 4 x^2 + 3 x + 1,\, x^3 +x^2 - 2x -1$), 
but we 
found similar results for other numbers.

\begin{figure}[!htb]
  \centering
   \begin{subfigure}{.9\textwidth}
     \centering
    \includegraphics[width=\textwidth]{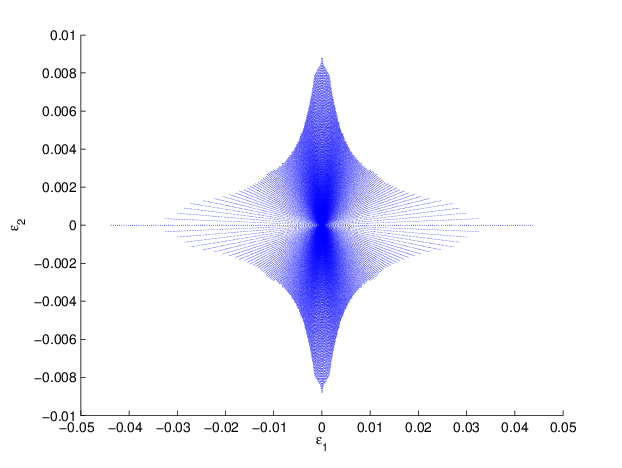}
    \caption{Domain of analyticity for Model \eqref{model1}.} \label{dom1}
  \end{subfigure}
   \begin{subfigure}{.9\textwidth}
     \centering
    \includegraphics[width=\textwidth]{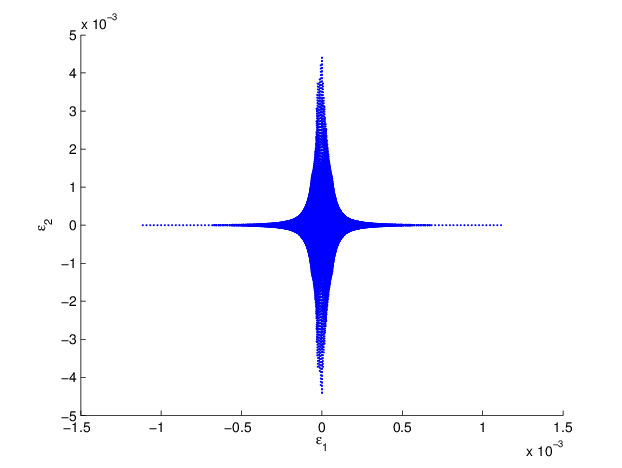}
    \caption{Domain of analyticity for Model \eqref{model2}.} \label{dom2}
  \end{subfigure}
  \caption{Values of $(\ep_1,\ep_2)$ where the models have analytic invariant tori.}
  \label{domains}
\end{figure}

For a fixed rotation, 
we use the algorithm to explore rays in parameter space $(\ep_1,\ep_2)$.
In Figure \ref{domains} we plot the 
domains of analyticity for Models \eqref{model1} and \eqref{model2}, that is,
values of $(\ep_1,\ep_2)$ where the 
algorithm converged.
The boundary of the domains of analyticity for each model 
appear to be smooth curves (at least
locally).  For Model \eqref{model1} the domain 
appears to have reflection symmetries over
both the $\ep_1$ and $\ep_2$ axes. Model \eqref{model2} 
appears to only have reflection
symmetry over the $\ep_1$ axis. 

\begin{figure}[!htb]
  \centering
   \begin{subfigure}{.7\textwidth}
     \centering
    \includegraphics[width=\textwidth]{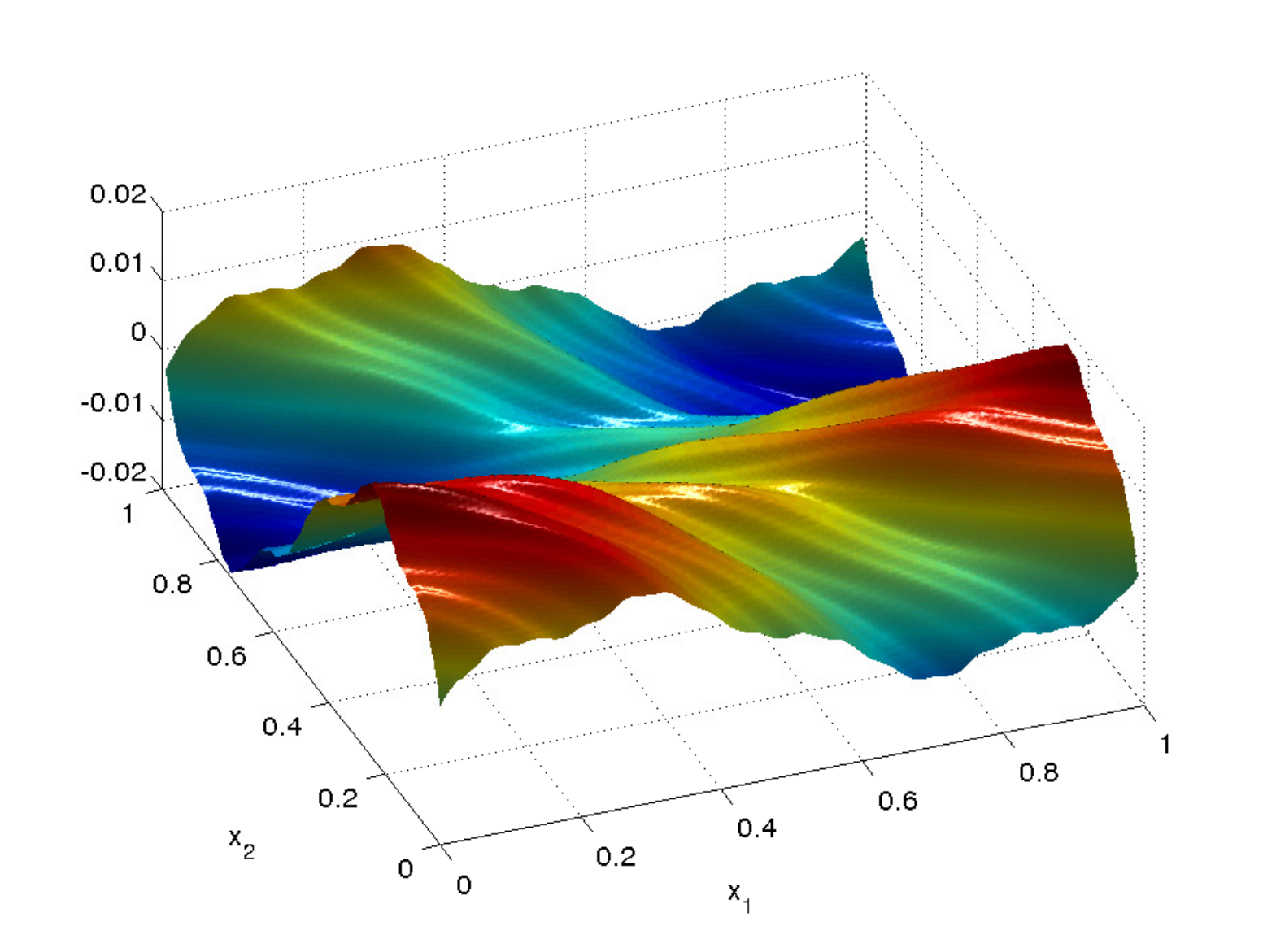}
    \caption{Torus near breakdown for Model \eqref{model1}.} \label{rid1}
  \end{subfigure}
   \begin{subfigure}{.7\textwidth}
     \centering
    \includegraphics[width=\textwidth]{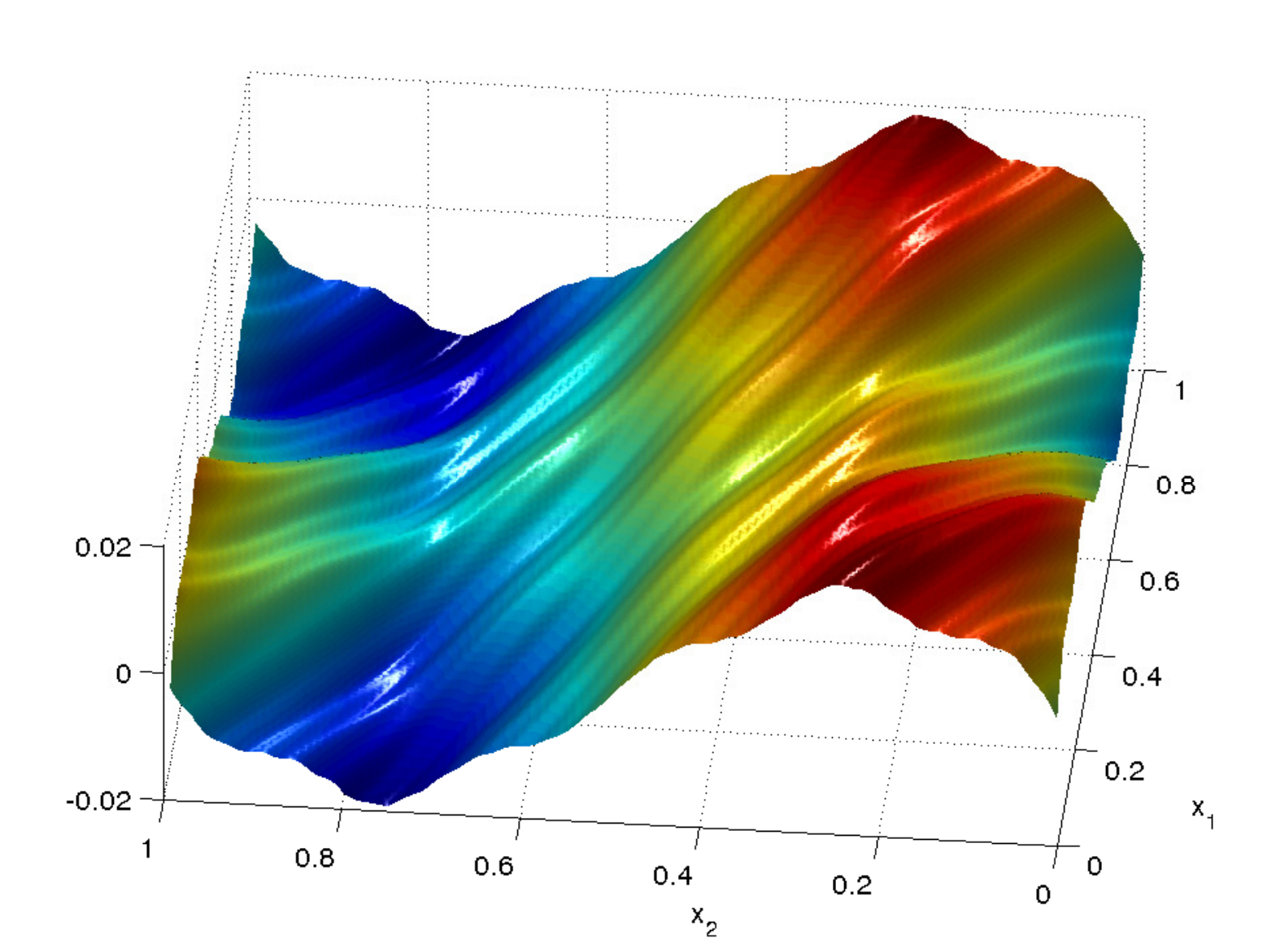}
    \caption{Rotated view of the torus.} \label{rid2}
  \end{subfigure}
  \caption{Two views of a torus near breakdown for Model \eqref{model1}.
Ridges appear as though $h$ has a frequency close to $\omega\alpha$.}
  \label{ridges}
\end{figure}

\begin{figure}[!htb]
  \centering
  \includegraphics[height=2.8in]{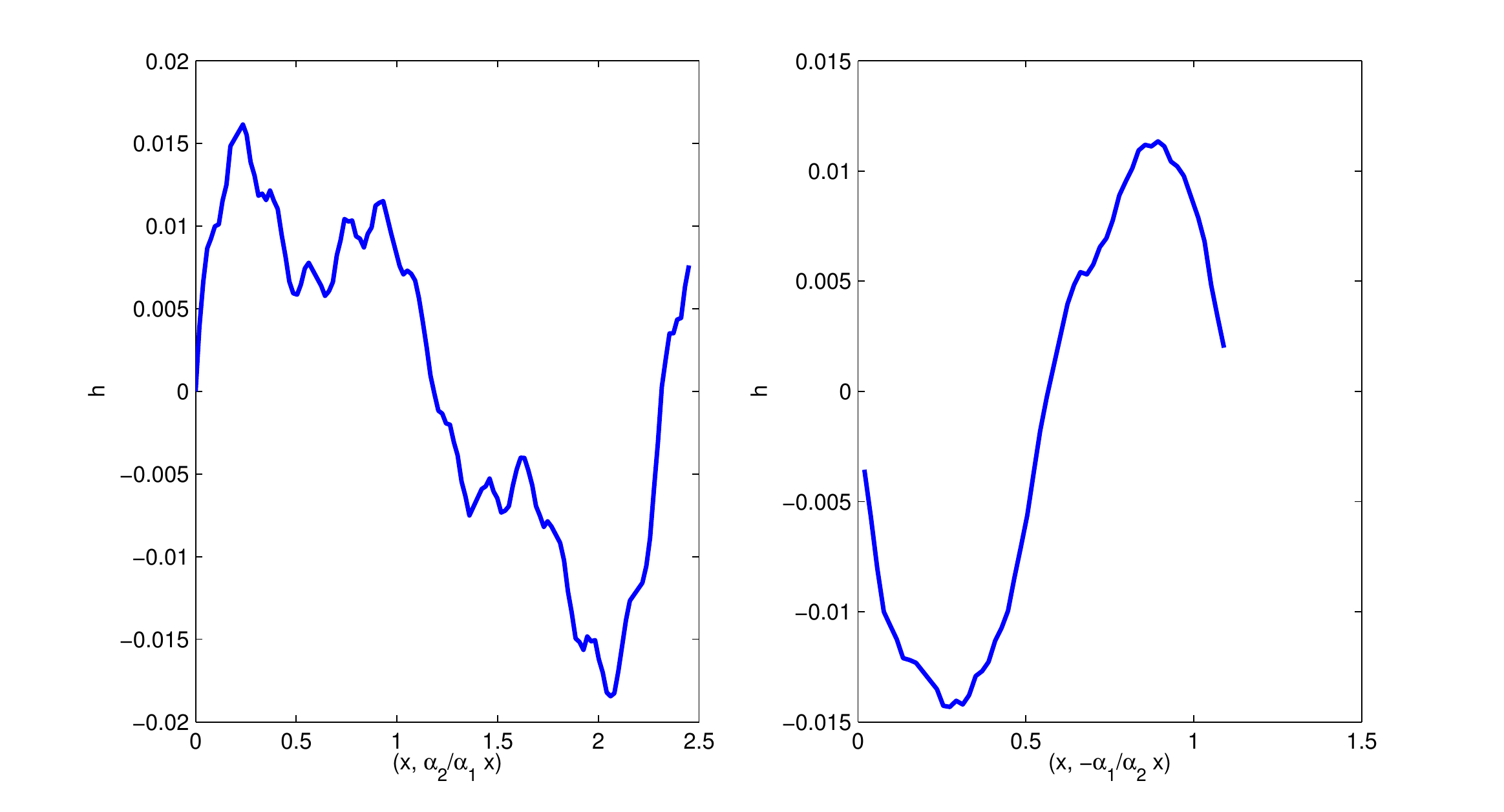}
  \caption{Left plot is of a cut along the direction $\omega \alpha$
of the torus from Fig. \ref{ridges}. The right plot is a surface cut
along the direction $\omega \alpha^\perp$. The torus is much smoother 
along the $\omega \alpha^\perp$ direction. }
  \label{cut}
\end{figure}

\begin{figure}[!htb]
  \centering
   \begin{subfigure}{.9\textwidth}
     \centering
    \includegraphics[width=\textwidth]{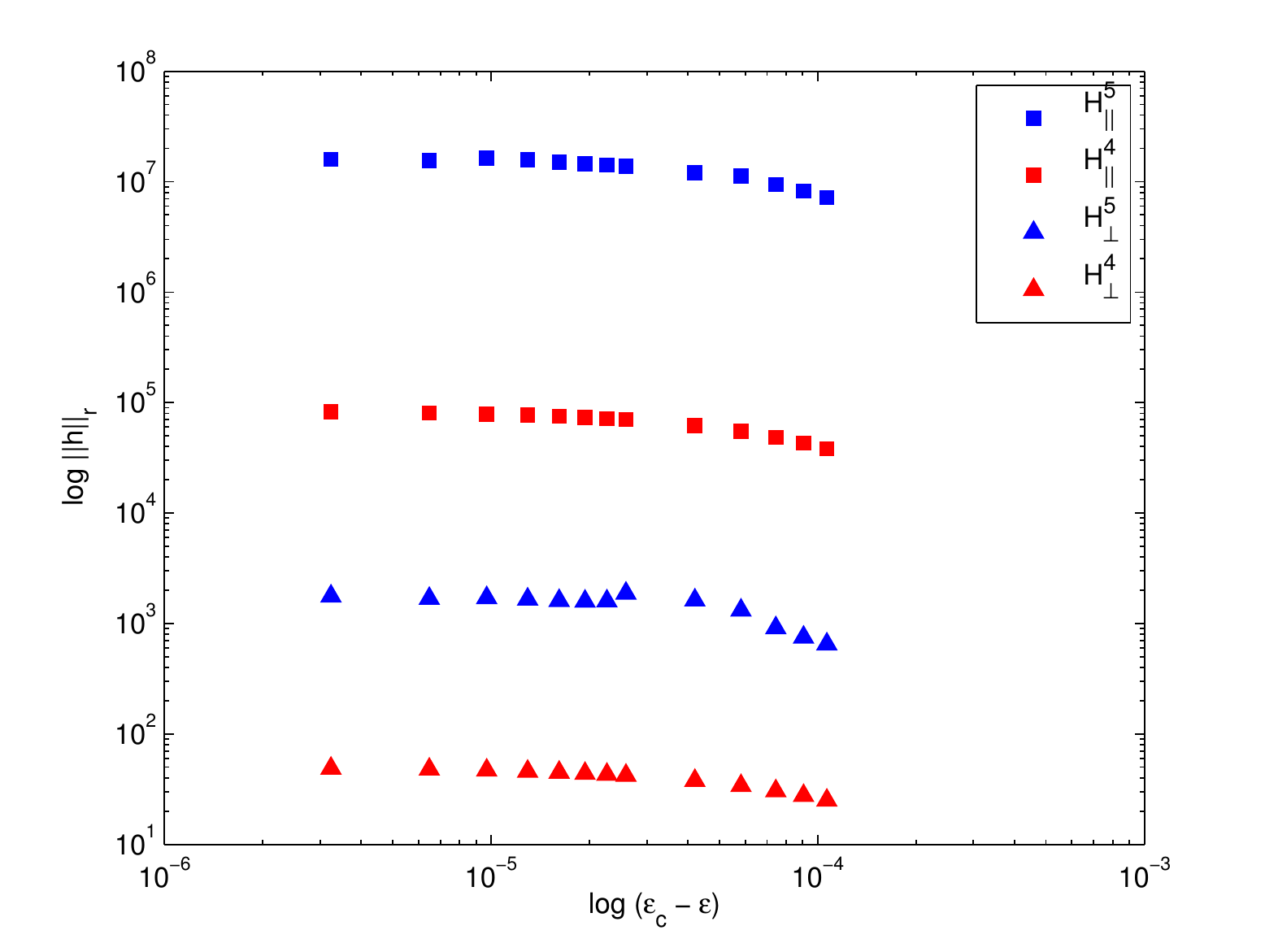}
    \caption{Log-log plots of $\|h\|_{r, \|}$ and $\|h\|_{r, \perp}$ vs. 
$\ep_{1,{\rm crit}}-\ep_1$ for $r=4,5$. Model \eqref{model1}.} \label{loglog1}
  \end{subfigure}
   \begin{subfigure}{.9\textwidth}
     \centering
    \includegraphics[width=\textwidth]{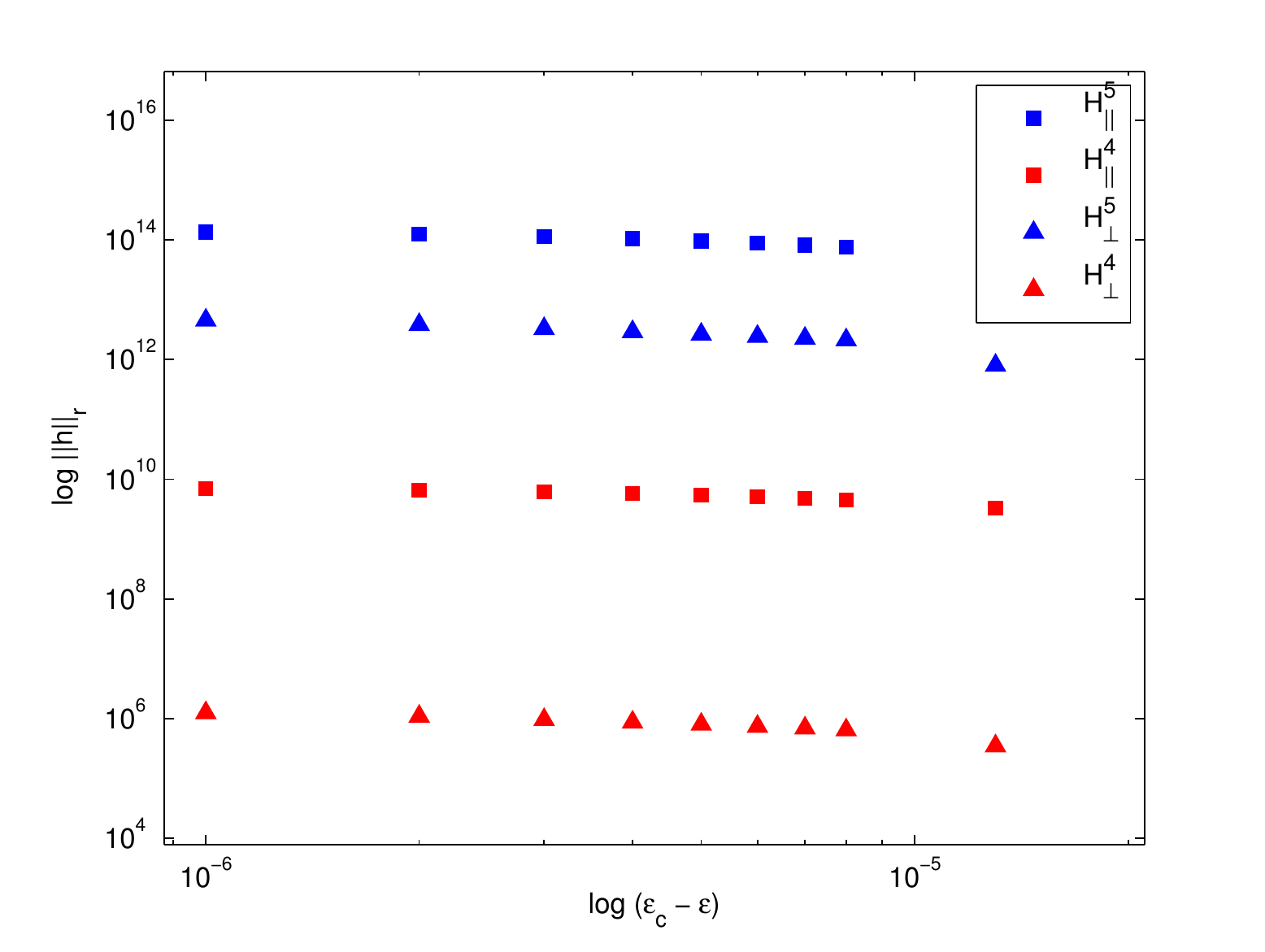}
    \caption{Log-log plots of $\|h\|_{r, \|}$, $\|h\|_{r, \perp}$ vs. 
$\ep_{1,{\rm crit}}-\ep_1$ for $r=4,5$. Model \eqref{model2}.} \label{loglog2}
  \end{subfigure}
  \caption{ $\|h\|_{r, \|}$ and $\|h\|_{r, \perp}$
blow up at different rates as $(\ep_1,\ep_2)$ increase along
the line $\ep_2 = \tan \left(\frac{\pi}{5}\right)\ep_1$ for 
Model \eqref{model1}. \ref{loglog1} shows blow-up behavior for Model \eqref{model1},
and \ref{loglog2} shows blow-up behavior for Model \eqref{model2}.
The $H^r_{\|}$-norms blow up much faster
than the $H^r_{\perp}$-norms.}
  \label{aniso}
\end{figure}

Figure \ref{ridges}
shows an invariant torus (from Model \eqref{model1})
for parameter values near the breakdown of analyticity (\ref{rid1} and
\ref{rid2} are two different views of this torus). One can see that
the torus oscillates rapidly in a single direction (this is the direction 
$\omega \alpha$). This ``filamented'' appearance indicates an anisotropic
breakdown: the derivatives of $h$ blow up faster in the direction
$\omega \alpha$ than they do in the direction $\omega \alpha^\perp$.
Figure \ref{cut} shows two curves that lie on the surface
depicted in Figure \ref{ridges}. These curves are obtained by cutting
the surface in Figure \ref{ridges} along the directions $\omega \alpha$ and
$\omega \alpha^\perp$. Along, $\omega \alpha$, we see fast oscillations
(left graph) compared to the cut along $\omega \alpha^\perp$ (right graph). 
(A more quantitative description of the anisotropic breakdown 
is given  in Figure \ref{aniso}, described below.)

\begin{figure}[!htb]
  \centering
  \includegraphics[height=3.0in]{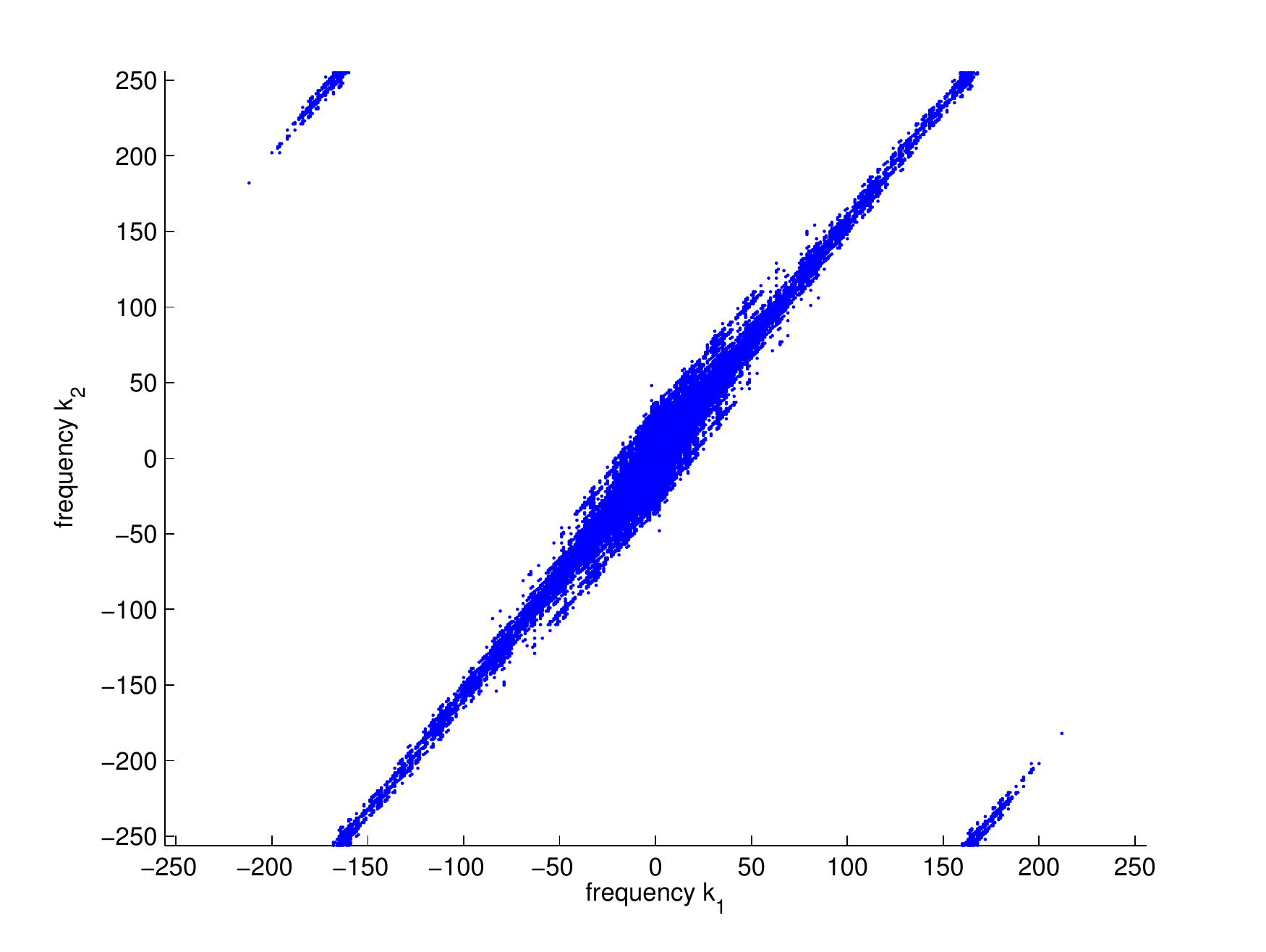}
  \caption{Fourier modes for a torus before breakdown for Model \eqref{model1}.
This is the same torus as in Figure \ref{ridges}.
A dot is plotted at the point $k=(k_1,k_2)$
if Fourier coefficient  $\hat{h}_{k}$ is non-zero.
Using $512^2$ coefficients, we have
$-256\leq k_1, k_2 \leq 255$.} 
  \label{fmodes}
\end{figure}

In Figure~\ref{fmodes} we have depicted the indices
$k$  for which the 
corresponding Fourier mode $\hat h_k$ of 
the torus in Figure~\ref{ridges}  is far from zero.
We observe that the support of the Fourier transform is 
clustered along a line in Fourier space. 

That means that the function $h$ can be aproximated by a 
function 
\[
h(\theta_1, \theta_2)   = \sum_{k \in \integer} 
\hat h_{k, \beta k}  e^{2 \pi i k \theta_1 + \beta k \theta_2 }
= H( \theta_1  + \beta \theta_2)
\]
which also lends credence to the belief that the result near
breakdown is basically a one dimensional function. 

Notice that, once we know that the final result is essentially a one 
dimensional function, it should be possible to perform linear 
changes of  the $\theta$ 
variables so that this line is approximately parallel to one of 
the coordinate axis. Then, it would be more efficient  to choose 
more Fourier modes along this direction. Of course, 
since our goal was to illustrate the appearence of this linear 
structure, we decided to use general algorithms that do 
not take advantage of its presence. 

In Figure \ref{aniso}, we plot a log-log scale of the
$H^r_\|$-norms
and $H^r_\perp$-norms (for $r=4,5$)
versus $\ep_{1,{\rm crit}}-\ep_1$, where $\ep_{1,{\rm crit}}$ is the critical
value of $\ep_1$ along the line $\ep_2 = \tan \left(\frac{\pi}{3}\right)\ep_1$.
We see that the $H_\|^r$-norms blow up much faster than
the $H^r_\perp$-norms, exhibiting the anisotropic breakdown.

\begin{figure}[!htb]
  \centering
   \begin{subfigure}{.9\textwidth}
     \centering
    \includegraphics[width=\textwidth]{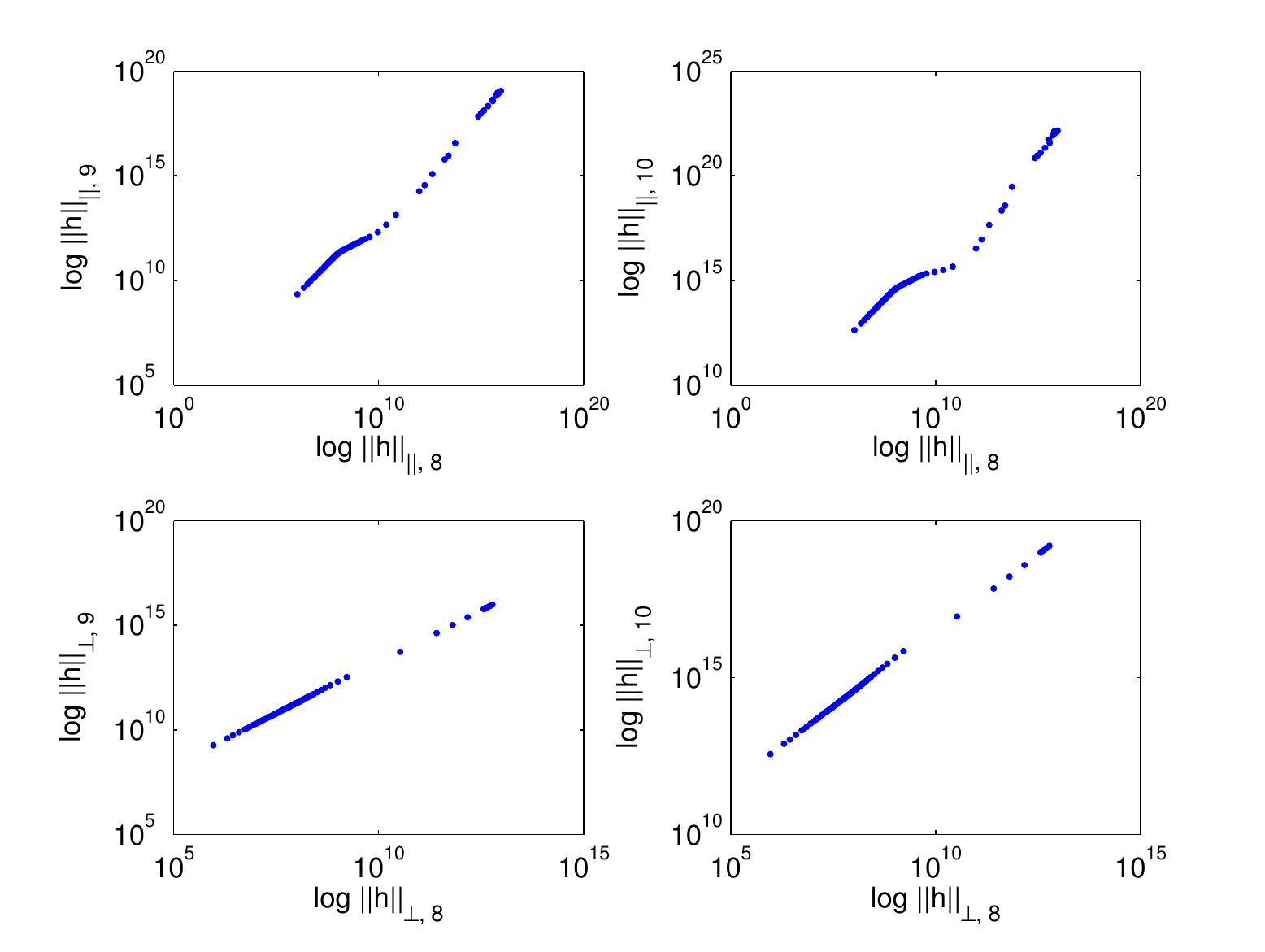}
    \caption{Log-log plots of $\|h\|_{r, \|}$ vs. $\|h\|_{8, \|}$ 
 and $\|h\|_{r, \perp}$ vs. $\|h\|_{8, \perp}$ for $r=9,10$.} \label{log_norms}
  \end{subfigure}
   \begin{subfigure}{.9\textwidth}
     \centering
    \includegraphics[width=\textwidth]{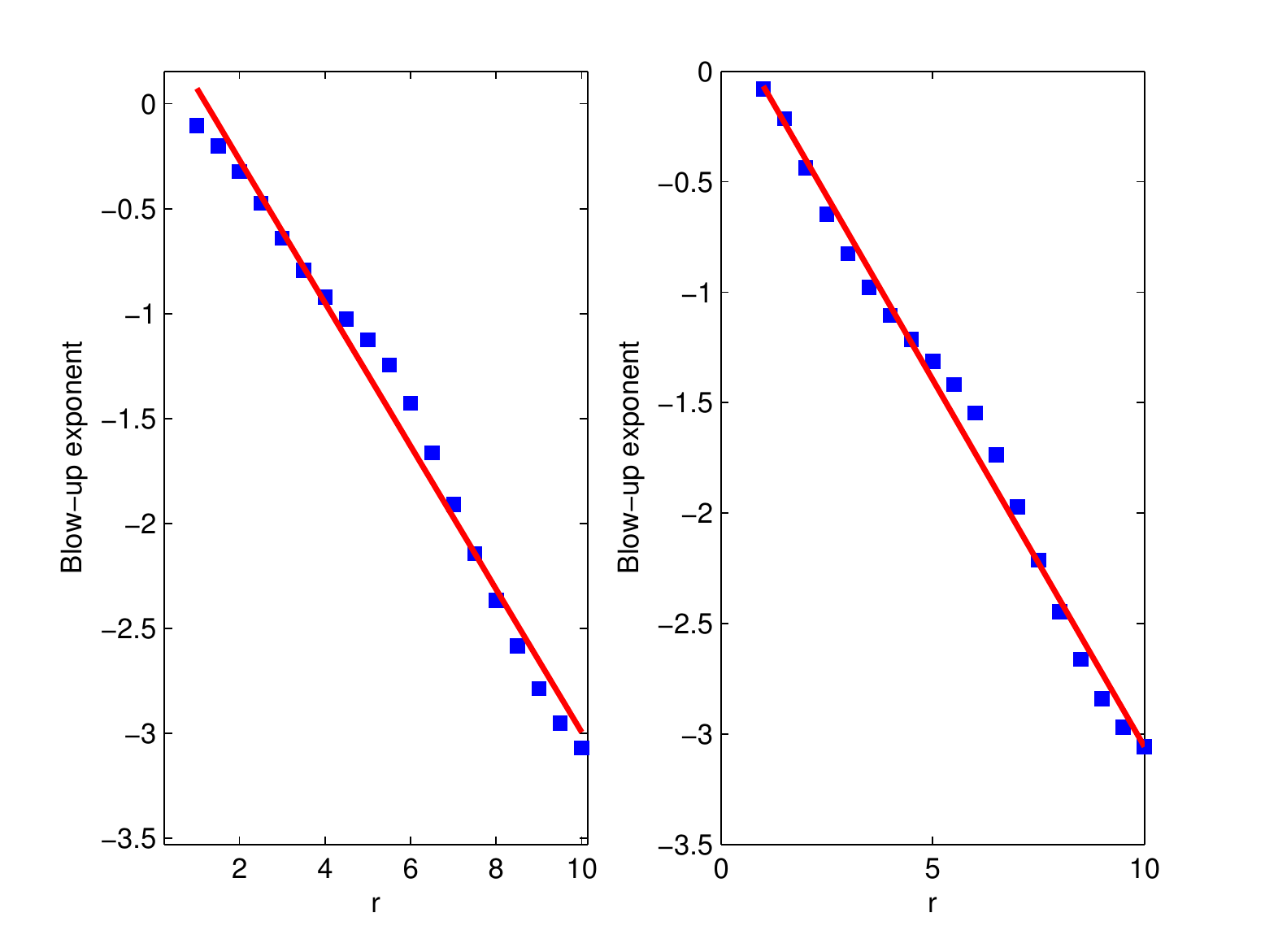}
    \caption{Blow-up exponents for  $\|h\|_{r,\|}$ vs. $r$.} \label{blowup_fit}
  \end{subfigure}
  \caption{\ref{log_norms} shows log-log plots of $H^r_\|$ 
vs $H^8_\|$ and $H^r_\perp$ vs $H^8_\perp$
for $r=9,10$.  (All plots
in this figure come from  tori for Model \eqref{model1}).
\ref{blowup_fit} shows the relationship between the
blow-up exponent of $\|h\|_{r,\|}$ as a function of  $r$
as $(\ep_1,\ep_2)$ move along the rays 
$\ep_2 = \tan \left(\frac{\pi}{5}\right)\ep_1$ and
$\ep_2 = \tan \left(\frac{\pi}{3}\right)\ep_1$. We
see that $\|h\|_{r,\|}\approx c (\ep_c - \ep)^{-\beta_\| r + \gamma_\|}$.
}
  \label{hnorms}
\end{figure}

In Figure \ref{log_norms}, we plot $\|h\|_{r,\|}$ against
$\|h\|_{8,\|}$, and $\|h\|_{r,\perp}$ against
$\|h\|_{8,\perp}$ for $r = 9, 10$ on log-log scales. As the
parameters $(\ep_1,\ep_2)$ approach critical values, the 
log-log scale behavior is linear. (The $H^r$-norm behaves
like the $H^r_\|$-norm because of the anisotropic behavior.)

The blow-up behavior seen in Figure \ref{aniso} fits a power-law 
form $\|h\|_{r,\|} \approx C (\ep_{\rm crit} - \ep)^{p(r)}$, with exponent
$p$ depending on the order of the Sobolev norm. Computing
the exponent for different values of $r$, we find a nearly linear
dependence on $r$. 
 In Figure \ref{blowup_fit} we plot the blow-up exponent for values
$r =1 ,1.5,2, \ldots , 10$
as computed along two different lines in parameter space
for Model \eqref{model1}. The blow-up exponent is
 $p(r) = -\beta_\| r + \gamma_\|$
for some constants $\beta_\| , \, \gamma_\|$. 
The fit-line
in the left plot has $\beta_\| \approx 0.34$ and $\gamma_\| \approx 0.41$,
while in right plot $\beta_\| \approx 0.33$ and $\gamma_\| \approx 0.26$.
With higher-accuracy numerics, we expect to have similar plots for
the $H^r_{\perp}$-norms, too. These subdominant norms are measured
with lower accuracy and more susceptible to noise in the numerics.
We can see that they blow up like a power as $(\ep_1, \ep_2)$
approach the boundary of analyticity
(Figure \ref{aniso}), but measuring the exponents
as functions of the regularity will require more powerful numerics.

%

\section{Conclusions} \label{sec:conclusions}
We have studied numerically the problem of breakdown of analyticity
of hull functions in quasi-periodic media. 
This transition corresponds to the physical effect of transition between 
pinned down or sliding ground states. 

Our computation has been 
carried out by efficient algorithms backed up by a-posteriori theorems that 
allow to compute confidently extremely close to the break up. 
We have uncovered certain regularities that, following the style of 
\cite{Greene79}, we call \emph{Assertions}. 

{\bf Assertion I} 
\emph{The boundary of the domain of analyticity is a smooth curve (at least locally in the region explored).}

Note that, following \cite{AubryA80} this is analogous to 
what happens in phase transitions, where 
the phase transitions -- which also correspond to 
a breakdown of analyticity -- happen often in 
smooth surfaces (Gibbs rule) \cite{Israel79}.
The fact that the transitions happen locally in a smooth manifolds
is one of the predictions of the renormalization group because they are the 
stable manifolds of a fixed point.

Of course, it can well happen that there
are regions where the phase transition has cusps or other singularities, but we have not 
encountered them.  It is, of course, quite conceivable that a more detailed exploration 
will uncover these singularities. As a cautionary tail,
we mention that the paper \cite{CallejaL10} 
uncovered singularities in the case of  twist mappings that appeared only when a very 
large range of parameters was explored. The singularities are an indication that the 
renormalization group dynamics may have a more complicated behavior. 
Note that the apparent cusps found in Figure \ref{domains} near the coordinate axis are not 
real singularities. The potential has symmetries that the reflections along axis 
with the change in the origin.

{\bf Assertion II } 
\emph{There are scaling behaviors near the blow up} 

We have seen that near the blow up we have 
\[
|| h_\ep ||_r  \approx  C  |\ep - \ep_c|^{- \beta r + \gamma}
\]
where $\beta$ (and perhaps  $\gamma$ too) is a universal number. 

Actually, we have found a more precise version of scalings in each direction. 

{\bf Assertion III}
\emph{The blow up of the norms is very anisotropic}

If we define 
\[
\begin{split}
& || h ||_{r, ||} = || (\omega\alpha \cdot \nabla)^r h ||_{L^2} \\
& || h ||_{r, \perp} = || (\omega\alpha^\perp \cdot \nabla)^r h ||_{L^2} \\
\end{split}
\]
we have  near the breakdown
\[
|| h_\ep ||_{r,||}  \approx  C | |\ep - \ep_c|^{- \beta_{||} r + \gamma_{||}}.
\]

Moreover, when $\ep \to \ep_c$, we have:
\[
|| h_\ep ||_{r, \perp} / || h_\ep||_{r, ||}  \to 0 .
\]

Note that, as a particular case of the assertion, the tori remain somewhat regular up
to the breakdown along lines. 

\begin{remark}
We think that it is also possible that there are scaling relations
\[
|| h_\ep ||_{r,\perp}  \approx  C | |\ep - \ep_c|^{- \beta_{\perp}+ \gamma_{\perp}}
\]
with smaller exponents than those for the parallel norm. 
Unfortunately, verifying the above scaling relations is beyond reach at the moment. 
The fact that the asymptotic expansion is subdominant makes it harder to compute 
numerically (the calculations are contaminated by the larger effect) and also 
one can expect that this effect will manifest itself only much closer to breakdown. 
\end{remark}

\section*{Acknowledgements} 
Part of this work was done while the authors were affiliated 
with Univ. of Texas. We thank S. Hernandez and X. Su for many 
discussions about this problem and about numerical issues. 
We also thank the Center for Nonlinear Analysis (NSF Grants 
No. DMS-
0405343 and DMS-0635983), where part of this research was carried out.
\bibliographystyle{alpha}
\bibliography{numerics}
\end{document}